# Deep-learning continuous gravitational waves: Multiple detectors and realistic noise

Christoph Dreissigacker[*] and Reinhard Prix

*Max Planck Institute for Gravitational Physics (Albert-Einstein-Institute), D-30167 Hannover, Germany
and Leibniz Universität Hannover, D-30167 Hannover, Germany*

The sensitivity of wide-parameter-space searches for continuous gravitational waves is limited by computational cost. Recently it was shown that deep neural networks (DNNs) can perform all-sky searches directly on (single-detector) strain data [C. Dreissigacker *et al.*, Phys. Rev. D **100**, 044009 (2019)], potentially providing a low-computing-cost search method that could lead to a better overall sensitivity. Here we expand on this study in two respects: (i) using (simulated) strain data from two detectors simultaneously, and (ii) training for directed (i.e., single sky-position) searches in addition to all-sky searches. For a data time span of $T = 10^5$ s, the all-sky two-detector DNN is about 7% less sensitive (in amplitude $h_0$) at low frequency ($f = 20$ Hz), and about 51% less sensitive at high frequency ($f = 1000$ Hz) compared to fully-coherent matched-filtering (using WEAVE). In the directed case the sensitivity gap compared to matched-filtering ranges from about 7%–14% at $f = 20$ Hz to about 37%–49% at $f = 1500$ Hz. Furthermore we assess the DNN's ability to generalize in signal frequency, spin down and sky-position, and we test its robustness to realistic data conditions, namely gaps in the data and using real LIGO detector noise. We find that the DNN performance is not adversely affected by gaps in the test data or by using a relatively undisturbed band of LIGO detector data instead of Gaussian noise. However, when using a more disturbed LIGO band for the tests, the DNN's detection performance is substantially degraded due to the increase in false alarms, as expected.

DOI: 10.1103/PhysRevD.102.022005

## I. INTRODUCTION

Observing gravitational waves from compact binary mergers has become routine [1–5]. The long-lasting but weak narrow-band signals from spinning non-axisymmetric neutron stars called continuous gravitational waves (CWs) however remain elusive at the current sensitivity of the Advanced LIGO [6] and Virgo [7] detectors. Despite great improvements in the search methods (see e.g., [8] for a recent review) and numerous searches conducted on past and recent detector data (see Refs. [9–11] for reviews), no CW discovery has been made yet.

The sensitivity of CW searches is typically limited by the prohibitive computing cost. A CW signal is expected to last longer than the observation time. Hence, to increase the signal-to-noise ratio (SNR) of a search, it needs to integrate over as much data as possible but for a typical fully coherent matched-filtering search the computing cost grows as a high power $\sim T^n$, $n \geq 5$ of the time span of data $T$. Therefore these statistically almost optimal searches [12] can only be performed with coherence times of days to weeks at most.

The main method to circumvent this limitation is to use semicoherent methods. These consist of using shorter coherent segments and combining their results incoherently resulting in an improved sensitivity at fixed computing cost [13,14]. Nevertheless the currently most sensitive wide parameter space searches (see e.g., [15–17]) are using massive amounts of computational resources, either in the form of local computer clusters or the distributed computing project Einstein@Home [18].

In this work we study the feasibility of *deep neural networks* (DNNs) as an alternative search method. DNNs have been shown to be able to approximate any Borel-measurable function [19,20] (see also [21] for a more general discussion). Therefore they should in principle be able to approximate gravitational-wave-search methods.

In fact the method of training a DNN, also called *deep learning*, has been established to be able to detect gravitational waves directly from strain data [22–28] for signals from mergers of compact objects. More recently it was used for the first time on simulated continuous gravitational wave signals [29] and it was applied to the related long transient signals [30]. Furthermore DNNs have been studied as a follow-up method for CW searches [31,32], as well as for parameter estimation of searches for compact binary merger signals [33,34] and for a multitude of other gravitational-wave-search related applications such as classifying disturbances (*glitches*) and searches for unmodeled burst signals [35–41].

---

[*]Corresponding author.
christoph.dreissigacker@aei.mpg.de

In this work we continue the effort toward building a competitive neural-network-based search method for CWs by gradually moving toward more realistic test- and training scenarios: by simultaneously using data from two detectors, by including directed search cases, and by testing the trained DNNs on Gaussian data with gaps and on real LIGO data with varying degrees of instrumental disturbances.

The plan of this paper is as follows: in Sec. II we define the new benchmark cases, we discuss the updated deep-learning approach in Sec. III, we characterize the performance of the DNNs on the benchmarks by testing them on Gaussian noise in Sec. IV and finally we extend this characterization to the intricacies of real detector noise in Sec. V. In Sec. VI we discuss our results.

## II. COMPARISON TEST BENCHMARKS

We characterize the DNNs as search method on the following two-detector benchmark searches, each assuming two different timespan baselines of $T = 10^5$ s and $T = 10^6$ s: an all-sky search and two directed searches pointing at the supernova remnants of Cassiopeia A (CasA) and G347.3-0.5 (G347), respectively. For the coherent matched-filter comparison we use the WEAVE search code [42] in the same way as in [29].

We measure the sensitivity of the DNNs and the matched-filtering searches by determining the detection probability $p_{\rm det}$ at a chosen false-alarm level of $p_{\rm fa} = 1\%$ per 50 mHz frequency band. The false-alarm level corresponds to a threshold on the respective detection statistic of the DNNs and the matched-filtering searches for a signal population of fixed signal amplitude given in terms of the sensitivity depth $\mathcal{D}$ [8,43], defined as

$$\mathcal{D} \equiv \frac{\sqrt{S_{\rm n}}}{h_0}, \qquad (1)$$

where $S_{\rm n}$ is the power spectral density of the noise at the signal frequency, and $h_0$ is the signal amplitude. In particular we will use $\mathcal{D}^{90\%}$ to refer to the 90%-upper limit depth, corresponding to a signal amplitude $h^{90\%}$ where a search method achieves a detection probability of $p_{\rm det} = 90\%$ at a false-alarm threshold of $p_{\rm fa} = 1\%$ per 50 mHz frequency band.

For reasons of speed and simplicity, at this stage of the project we still use simulated Gaussian noise for the DNN training and for the matched-filtering comparison. However, in Sec. V we do show tests of our DNN search pipeline on data with realistic gaps and also using real LIGO detector data.

TABLE I. Definition of all-sky (two-detector) benchmark searches.

| | |
|---|---|
| Data span | $T = 10^5$ s/$T = 10^6$ s |
| Detectors | LIGO Hanford (H1) + Livingston (L1) |
| Noise | Stationary white Gaussian[a] |
| Sky-region | All-sky |
| Frequency band | $f \in [20, 1000]$ Hz |
| Spin-down range | $\dot{f} \in [-10^{-10}, 0]$ Hz/s |

[a]Excluding the real-data tests in Sec. V B

The two-detector benchmarks are similar to the previously-used single-detector benchmarks of [29], as they encompass data spans of $T = 10^5$ s and $T = 10^6$ s, and the all-sky searches cover the same parameter space as the previous single-detector all-sky cases (see Table I).

The new directed search benchmarks are derived from the Einstein@Home multidirected search for CWs in O1 data (cf. [16]). They cover a frequency range of 20–1500 Hz and large ranges of first and second-order spin-down, which are functions of the characteristic age of the targeted supernova remnant and the frequency (see Table II). Compared to the original search, however, the total observation time is substantially reduced to the two benchmark spans of $T = 10^5$ s and $T = 10^6$ s.

Similarly to [29], we limit the required matched-filtering computing cost by only searching a narrow frequency band of $\Delta f = 50$ mHz at a few representative starting frequencies in the range of 20–1500 Hz. The characteristics of the matched-filtering searches can be found in Table III.

## III. DEEP-LEARNING CWs

The approach used here is an evolved version of our previous deep-learning study in [29]: We train a noise-versus-signal classifier on strain data from two detectors. The input is provided as two separate channels per detector, each containing respectively the real and imaginary part of the Fourier transform of the strain data. This results in four input channels for our two-detector cases.[1]

TABLE II. Definition of directed benchmark searches, modeled after [16].

| | |
|---|---|
| Data span | $T = 10^5$ s/$T = 10^6$ s |
| Detectors | LIGO Hanford (H1) + Livingston (L1) |
| Noise | Stationary white Gaussian[a] |
| Sky-position | CasA/G347 |
| Frequency band | $f \in [20, 1500]$ Hz |
| Spin-down range | $-f/\tau \leq \dot{f} \leq 0$ Hz/s |
| Second order spin-down | $0$ Hz/$s^2 \leq \ddot{f} \leq 5f/\tau^2$ |
| Characteristic age $\tau$ | CasA: 330 yrs, G347: 1600 yrs |

[a]Excluding the real-data tests in Sec. V B

TABLE III. WEAVE coherent matched-filtering search parameters and characteristics.

| Search | Mismatch | Mean SNR loss | Templates |
|---|---|---|---|
| All-sky $T = 10^5$ s | 0.1 | 4% | $7 \times 10^{11}$ |
| All-sky $T = 10^6$ s | 0.2 | 8% | $4 \times 10^{14}$ |
| G347 $T = 10^5$ s | 0.1 | 5% | $1 \times 10^{10}$ |
| G347 $T = 10^6$ s | 0.2 | 10% | $6 \times 10^{12}$ |
| CasA $T = 10^5$ s | 0.1 | 5% | $6 \times 10^{10}$ |
| CasA $T = 10^6$ s | 0.2 | 10% | $3 \times 10^{13}$ |

However the networks could easily be trained for data from more detectors by adding additional channels. As for matched-filtering the additional data would increase the computational cost. Due to this straight forward generalization we only consider two-detectors as most matched-filtering searches at the moment only consider data from the two most sensitive detectors (e.g., see [8]).

We determine the maximal width in frequency of the signals in the allowed parameter space of the search and set the DNN input size to twice this width. This allows us to slide half overlapping windows over the frequency band to guarantee any signal is fully contained in at least one window. This leads to an increase of the DNN input size with observation time as well as with the number of detectors.

### A. Finding a network architecture

We started experimenting with the modified 1D-ResNet architecture from [29] and other 1D-versions of architectures like *InceptionResNet-v2* [44] which have proven successful for image recognition. For various different architectures we trained a network on a smaller number of samples for the $T = 10^5$ s, $f_0 = 1000$ Hz benchmark case. We compare the different networks' performance by calculating their detection probability on the validation set described in Sec. III B.

The architecture with the best detection probability in these experiments is an Inception-ResNet architecture: The InceptionResNet-v2 architecture was modified to feature one-dimensional inputs. For the $T = 10^5$ s cases this network was further enlarged by increasing the number of block repetitions by 2, in width by increasing the filter sizes by 2 as well as the number of filters in every convolutional layer by a factor of 4. The resulting network needs too much memory[2] for the larger inputs of the $T = 10^6$ s cases, therefore we use the original nonenlarged network for the $T = 10^6$ s benchmark cases.

The DNN input is first normalized by its standard deviation. The DNN output is created with a global average pooling layer and a dense layer with two final neurons and a softmax activation. The two outputs are encoding the estimated probabilities that the input contains a signal in noise $p_\text{signal}$ or pure noise, $p_\text{noise} = 1 - p_\text{signal}$, respectively.

The DNN was implemented in TENSORFLOW 2.0 [45] with its inbuilt Keras API (tf.keras). The CW signal generation was performed using the PYTHON SWIG-wrapping [46] of LALSUITE [47].

### B. DNN training and validation

We trained a total of 25 networks, one for each case listed in Table IV and in addition one all-sky and two directed networks trained for the entire respective search frequency range of the $T = 10^5$ s second searches. As in [29] each of the networks is trained with a set of synthesized input vectors, where half contain pure Gaussian noise, and half contain a signal added to the noise. The training set is built from 5 000 precomputed signals which are added to 24 dynamically generated noise realizations each. The noise realizations are also added as pure noise examples giving 240 000 samples in the training set in total.

The number of 5 000 signals was determined as a compromise between requirements in computing resources and the diminishing improvements which could be gained with a bigger training set (cf. [29] for details).

The signals are scaled to an evolving depth $\mathcal{D}_\text{training}(\text{epoch})$ which starts low, i.e., with louder signals, and then increases every five epochs until it reaches the final training depth $\mathcal{D}_\text{training}$, according to the following curriculum:

TABLE IV. Sensitivity depths $\mathcal{D}_\text{MF}^{90\%}$ achieved by the WEAVE coherent matched-filtering search for the (two-detector) all-sky (a-s) and directed cases defined in Tables I and II. The all-sky sensitivity is improved by a factor of approximately $\sqrt{2}$ compared to the single-detector values reported in [29], as expected for coherent matched filtering. As training the CasA case $T = 10^6$ s, $f_0 = 1500$ Hz required more GPU memory than available to us, we reduced the maximum frequency in the search to 1000 Hz.

| $\mathcal{D}_\text{MF}^{90\%}[\text{Hz}^{-1/2}]$ | 20 Hz | 100 Hz | 200 Hz | 500 Hz | 1000 Hz |
|---|---|---|---|---|---|
| a-s $T = 10^5$ s | 16.0 | 15.0 | 14.5 | 14.2 | 13.6 |
| a-s $T = 10^6$ s | 42.0 | 40.1 | 39.4 | 38.3 | 35.9 |

| $\mathcal{D}_\text{MF}^{90\%}[\text{Hz}^{-1/2}]$ | 20 Hz | 500 Hz | 1000 Hz | 1500 Hz |
|---|---|---|---|---|
| G347 $T = 10^5$ s | 18.5 | 17.1 | ⋯ | 16.5 |
| G347 $T = 10^6$ s | 46.1 | 43.9 | ⋯ | 43.1 |
| CasA $T = 10^5$ s | 17.5 | 16.3 | ⋯ | 15.7 |
| CasA $T = 10^6$ s | 46.1 | 43.8 | 43.4 | ⋯ |

---

[1]Note that a neural network with one input dimension and multiple channels is still commonly referred to as a 1D network.
[2]The largest GPU used (NVIDIA Quadro GV100) has 32 GB of GPU memory.

$$\mathcal{D}_{\text{training}}(\text{epoch}) = \frac{\mathcal{D}_{\text{training}}}{\gamma(\text{epoch mod 5})}, \quad (2)$$

where $\gamma(n) = (1.75, 1.5, 1.25, 1, 1, 1, \ldots)$, i.e., the signals get weaker until, after 15 epochs, the sensitivity depth reaches $\mathcal{D}_{\text{training}}$, which is chosen as the semianalytic estimate for the WEAVE-sensitivity depth $\mathcal{D}_{\text{MF}}^{90\%}$, using the method of Ref. [8]. At the time of training the final measured WEAVE matched-filtering sensitivity depths of Table IV were not yet available, which is why we used the faster (but less accurate) sensitivity-estimation instead.

This type of curriculum learning [48] is necessary to teach the network to find the weak signals at the final depth. Without it the network seemed unable to pick up the weak signals at the beginning and therefore was unable to learn at all. This is probably a consequence of the vastly increased number of parameters in the network compared to the network used in [29].

The DNNs were trained with a categorical cross entropy loss and the Adadelta optimizer [49]. They were each trained for 10 days on NVIDIA GPUs (RTX2060,70, 80(Ti), GV100, GTX 1660(Ti)) contained in the ATLAS computing cluster. By that time all the networks were fully trained, i.e., they did not show any significant improvement over the last couple of epochs.

During training we perform a validation step every five epochs where the detection probability is calculated on 20 000 independently-generated samples: 10 000 pure noise samples and 10 000 samples containing signals in Gaussian noise of the fixed depth $\mathcal{D}_{\text{training}}$.

In order to avoid a numerical overflow in the final softmax activation layer, we do not use the estimated softmax probabilities as a detection statistic. Instead we directly use the final linear network output which corresponds to $p_{\text{signal}}$ (i.e., the respective input to the final softmax activation) as a detection statistic. The detection probability is calculated in the usual way as the fraction of signal cases where this statistic crosses over the $p_{\text{fa}} = 1\%$ threshold.

## IV. CHARACTERIZING DNN PERFORMANCE ON GAUSSIAN NOISE

As the networks' parameters have been optimized for the training set and the network architecture (or *hyperparameters*) was optimized for the validation set, we need to evaluate the network's performance on an independent *test set* to fully characterize its performance as a CW detection method. This test set consists of noise and signals with randomly drawn parameters from a distribution isotropic in the sky and uniform in the other parameters. It is generated on-the-fly using the LALSUITE software library [46,47].

### A. Detection probabilities at fixed false alarm

The results in the following are presented in two ways:

TABLE V. Network sensitivity depths $\mathcal{D}_{\text{DNN}}^{90\%}$ for the (two-detector) all-sky (a-s) and directed search cases. The corresponding matched-filtering sensitivity depths are given in Table IV. As training the CasA case $T = 10^6$ s, $f_0 = 1500$ Hz required more GPU memory than available to us, we reduced the maximum frequency in the search to 1000 Hz.

| $\mathcal{D}_{\text{DNN}}^{90\%}[\text{Hz}^{-1/2}]$ | 20 Hz | 100 Hz | 200 Hz | 500 Hz | 1000 Hz |
|---|---|---|---|---|---|
| a-s $T = 10^5$ s | 14.9 | 13.2 | 12.4 | 10.6 | [a]9.0 |
| a-s $T = 10^6$ s | 29.6 | 17.5 | 13.9 | 9.7 | 7.9 |

| $\mathcal{D}_{\text{DNN}}^{90\%}[\text{Hz}^{-1/2}]$ | 20 Hz | 500 Hz | 1000 Hz | 1500 Hz |
|---|---|---|---|---|
| G347 $T = 10^5$ s | 16.3 | 13.6 | ⋯ | 11.1 |
| G347 $T = 10^6$ s | 33.9 | 11.7 | ⋯ | 1.3 |
| CasA $T = 10^5$ s | 16.4 | 13.4 | ⋯ | 11.5 |
| CasA $T = 10^6$ s | 28.1 | 0.0[b] | 1.4 | ⋯ |

[a]The given result is from a network trained on the whole frequency range, the specialized network performed worse, having a sensitivity depth of 7.9 Hz$^{-1/2}$ (see Fig. 1).
[b]The network did not reach 90% detection probability even at the lowest depth tested $\mathcal{D}^{90\%} = 0.1$ Hz$^{-1/2}$

(1) the detection probability $p_{\text{det}}^{\text{DNN}}$ obtained at false alarm $p_{\text{fa}} = 1\%$ per 50 mHz frequency band for a signal population of fixed depth $\mathcal{D}_{\text{MF}}^{90\%}$, for which the coherent WEAVE matched-filtering search achieves $p_{\text{det}} = 90\%$.
(2) The "upper-limit" depth $\mathcal{D}_{\text{DNN}}^{90\%}$ for the network, where it achieves a detection probability of $p_{\text{det}}^{\text{DNN}} = 90\%$ at $p_{\text{fa}} = 1\%$ per 50 mHz frequency band.

The measured DNN sensitivity on the all-sky search benchmarks is given in Tables V and VI. Similar to the previous single-detector results in [29], for $T = 10^5$ s at low frequencies the DNN achieves a performance close to matched filtering, while it increasingly falls behind for higher frequencies. However, for the $T = 10^6$ s cases the new network does not perform well and quickly drops to low sensitivity at increasing frequency.

The measured DNN sensitivity for the directed search benchmarks is also given in Tables V and VI. The results are similar in nature to the all-sky search results. For the $T = 10^5$ s searches for both targets the DNN gets relatively close to the matched-filtering performance, while for the $T = 10^6$ s searches they rapidly lose sensitivity when going to higher frequencies.

Note that in the $T = 10^6$ s searches our new network seems to perform worse and fall off more rapidly compared to the previous benchmark results in [29]. This loss in performance at $T = 10^6$ s can be traced back to two reasons: First, the new network architecture was optimized only for the $T = 10^5$ s searches, and second we only trained a single network instance instead of picking the best from an ensemble of 100 networks, as was done in [29], due to the increased hardware requirements of the new network architecture.

TABLE VI. Network detection probabilities $p_{det}^{DNN}$ with 95% error region for the (two-detector) all-sky (a-s) cases and directed cases for signals at the matched-filtering sensitivity depths $\mathcal{D}_{MF}^{90\%}$ given in Table IV. As training the CasA case $T = 10^6$ s, $f_0 = 1500$ Hz required more GPU memory than available to us, we reduced the maximum frequency in the search to 1000 Hz.

| $p_{det}^{DNN}[\%]$ | 20 Hz | 100 Hz | 200 Hz | 500 Hz | 1000 Hz |
|---|---|---|---|---|---|
| a-s $T = 10^5$ s | $84.4^{+4.0}_{-2.3}$ | $79.5^{+3.3}_{-3.5}$ | $78.1^{+3.3}_{-2.9}$ | $70.4^{+3.3}_{-3.4}$ | [a]$59.1^{+4.4}_{-3.7}$ |
| a-s $T = 10^6$ s | $60.5^{+3.7}_{-3.1}$ | $24.5^{+3.1}_{-3.1}$ | $11.2^{+3.1}_{-2.4}$ | $3.3^{+2.4}_{-1.3}$ | $0.7^{+0.7}_{-0.8}$ |

| $p_{det}^{DNN}[\%]$ | 20 Hz | 500 Hz | 1000 Hz | 1500 Hz |
|---|---|---|---|---|
| G347 $T = 10^5$ s | $79.6^{+3.1}_{-3.1}$ | $71.8^{+5.1}_{-7.7}$ | ⋯ | $64.2^{+3.6}_{-3.6}$ |
| G347 $T = 10^6$ s | $71.2^{+3.1}_{-3.0}$ | $2.6^{+2.1}_{-1.2}$ | ⋯ | $0.4^{+1.1}_{-0.6}$ |
| CasA $T = 10^5$ s | $86.4^{+3.3}_{-5.5}$ | $75.2^{+3.1}_{-4.4}$ | ⋯ | $65.5^{+3.4}_{-3.6}$ |
| CasA $T = 10^6$ s | $54.6^{+3.3}_{-3.7}$ | $0.6^{+0.6}_{-0.7}$ | $0.7^{+1.0}_{-0.7}$ | ⋯ |

[a]The given result is from a network trained on the whole frequency range, the specialized network performed worse, reaching a detection probability of $47.9^{+4.0}_{-3.8}\%$ (see Fig. 1).

### B. Generalization

One of most promising features of the DNN benchmarks results found in [29] was the surprising capability of the DNN to generalize to signal parameters it was not trained for. We confirm this feature for the new DNN used for the $T = 10^5$ s all-sky searches in this work for frequency, signal strength, spin-downs and sky position. For the $T = 10^5$ s directed search benchmarks introduced in this work, we also find a remarkable capability to generalize despite being less general than the all-sky DNNs. Given the rather poor DNN performance on the $T = 10^6$ s cases, discussed in Sec. IV A, we do not include those in the generalization tests shown here.

#### 1. Frequency

To avoid large computational cost for the training, we want to use as few networks as possible, optimally even a single one, to cover the search band with a reasonable sensitivity. Therefore we want to compare how a network trained over the full frequency band compares to "specialized" narrow-band networks trained on 50 mHz bands, when tested over the full frequency range.

The results of these tests for the all-sky two detector $T = 10^5$ s search can be found in Fig. 1. We find that the "specialized" networks trained for small frequency bands generalize well to lower frequency and slightly worse but still quite well to higher frequencies, confirming the findings in the single-detector case in [29]. However, the network trained over the full frequency range shows promise as it seems to fall only marginally behind the specialized networks for most frequencies—even beating

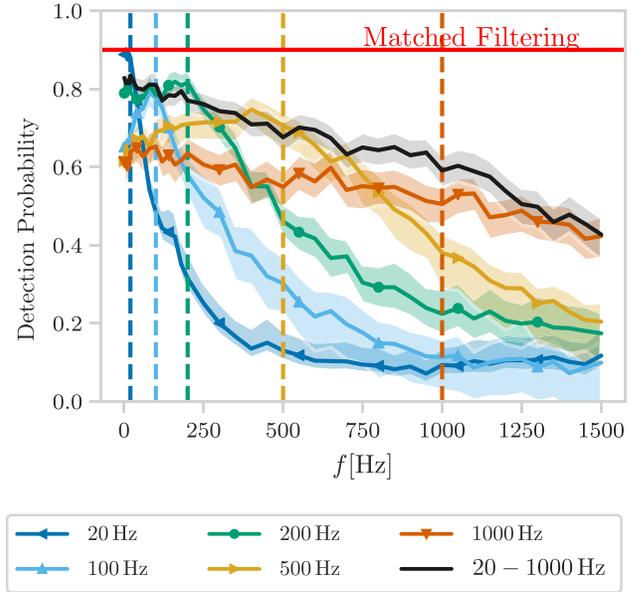

FIG. 1. Detection probability $p_{det}$ versus injection frequency $f$ for the all-sky networks trained at five different frequencies and for a network trained with signals drawn from the full frequency range (solid black line). The dashed vertical lines mark the respective training frequencies for the five "specialized" networks. The solid red horizontal line represents the coherent matched-filtering detection performance. The shaded areas around each curve show the 95% error regions. The analogous single-detector result is found in Fig. 6(a) of [29].

some of the specialized the networks at their training frequencies.

In the case of directed-search DNNs shown in Fig. 2, we see much narrower generalization around the trained frequencies of the "specialized" networks compared to the all-sky cases. The better generalization of the all-sky networks is likely due to the (known) near-degeneracy between frequency and sky position for short observation times. The networks trained over the full frequency in the directed cases significantly fall behind the specialized networks at their respective frequencies, contrary to our finding in the all-sky case. This is also likely connected to the mentioned near-degeneracy, as in the directed case, increasing the frequency range forces the network to learn many new signal shapes, while in the all-sky case the new signal shapes where already covered via signals from different sky-positions.

#### 2. Signal strength

To fully characterize a search method it is important to look at the detection efficiency curve, i.e., the detection probability for different signal strengths, shown in Fig. 3. This is especially interesting given that we use a single (final) depth $\mathcal{D}_{training}$ for training (cf. Sec. III B). The observed efficiency curves are very similar across the different searches, hence we only show two representative examples, the directed CasA search at $T = 10^5$ s, $f_0 = 20$ Hz, and an all-sky search at $T = 10^5$ s, $f_0 = 1000$ Hz.

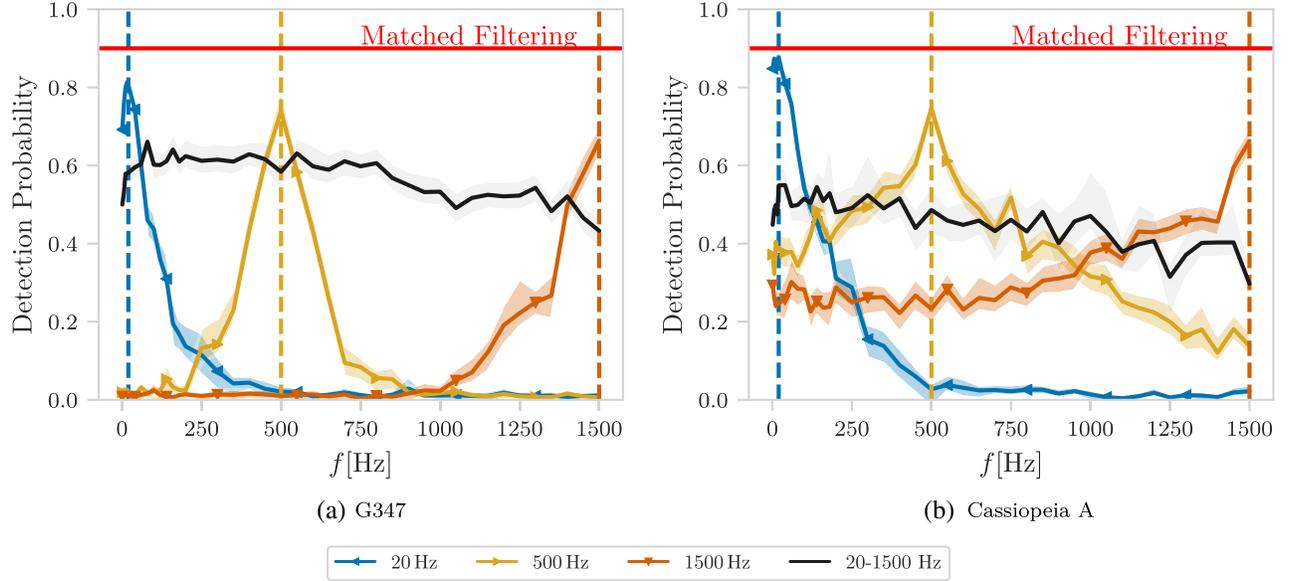

FIG. 2. Detection probability $p_{\text{det}}$ versus injection frequency $f$ for networks trained at three different frequencies for the CasA and the G347 target, respectively, and for a network trained with signals drawn from the full frequency range (solid black line). The dashed vertical lines mark the respective training frequencies for the three networks. The solid red horizontal line represents the coherent matched-filtering detection performance. The shaded areas around each curve show the 95% error regions.

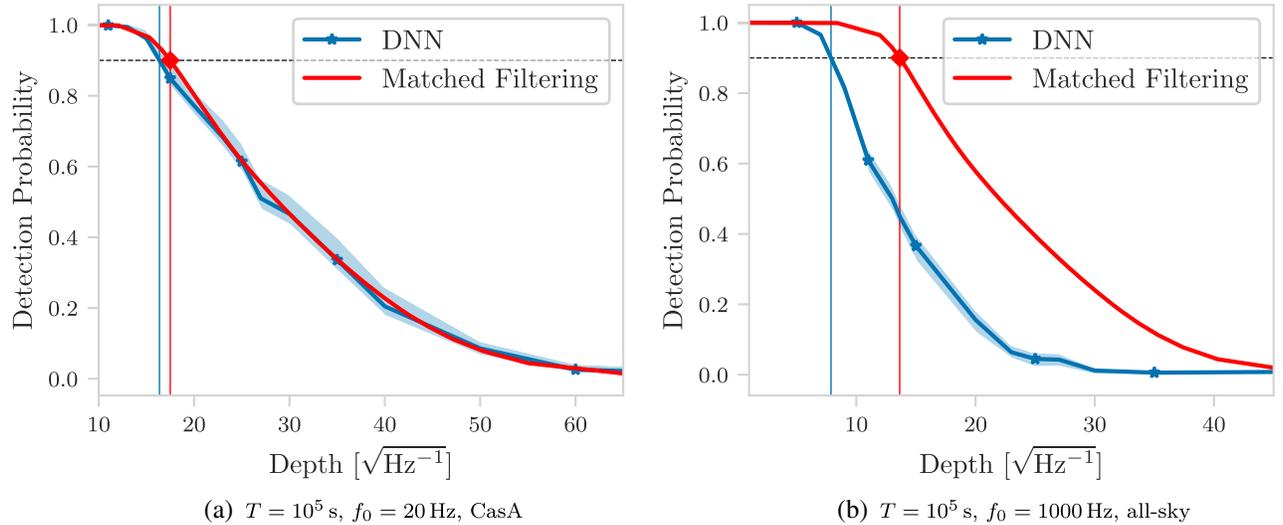

FIG. 3. Detection probability $p_{\text{det}}$ versus injection depth $\mathcal{D}$ for networks trained on the respective matched-filtering depth $\mathcal{D}_{\text{MF}}^{90\%}$ (indicated by the vertical solid line with the diamond at 90%). The second vertical line which crosses the DNN curve at 90% gives the sensitivity depth for the DNN at 90% detection probability. The shaded region around the DNN curve is the 95% error region. The respective errors for the matched-filtering results are smaller than the thickness of the curve.

In general the DNNs show qualitatively similar efficiency curves as matched-filter searches. We notice especially that the curves become almost indistinguishable for the low frequency cases while for higher frequency the DNNs relations seems to be shifted toward their overall lower sensitivity.

#### 3. Spin-downs

Another important aspect we want to consider is the detectability of signals with first and second order spin-downs outside of the training range, shown in Figs. 4 and 5, respectively.

For the all-sky searches we observe a similar behavior (not shown) in $\dot{f}$ to the results reported in Fig. 7 of [29] for the single-detector benchmarks: a plateau of nearly constant detection probability by far exceeding the training region. For the directed searches, however, we see a different behavior in $\dot{f}$, shown in Fig. 4: the DNNs plateau of nearly-constant detection probability falls off starting from the maximum absolute spin-down value of the training set. The generalization is not completely symmetric, though, and extends to larger negative than positive spin-downs. This might be an effect of the purely positive second-order spin-down breaking the degeneracy. The strong generalization beyond trained spin-downs of the all-sky DNNs might be due to the (known) near-degeneracy between spin-down and sky position for short observation times compared to a year.

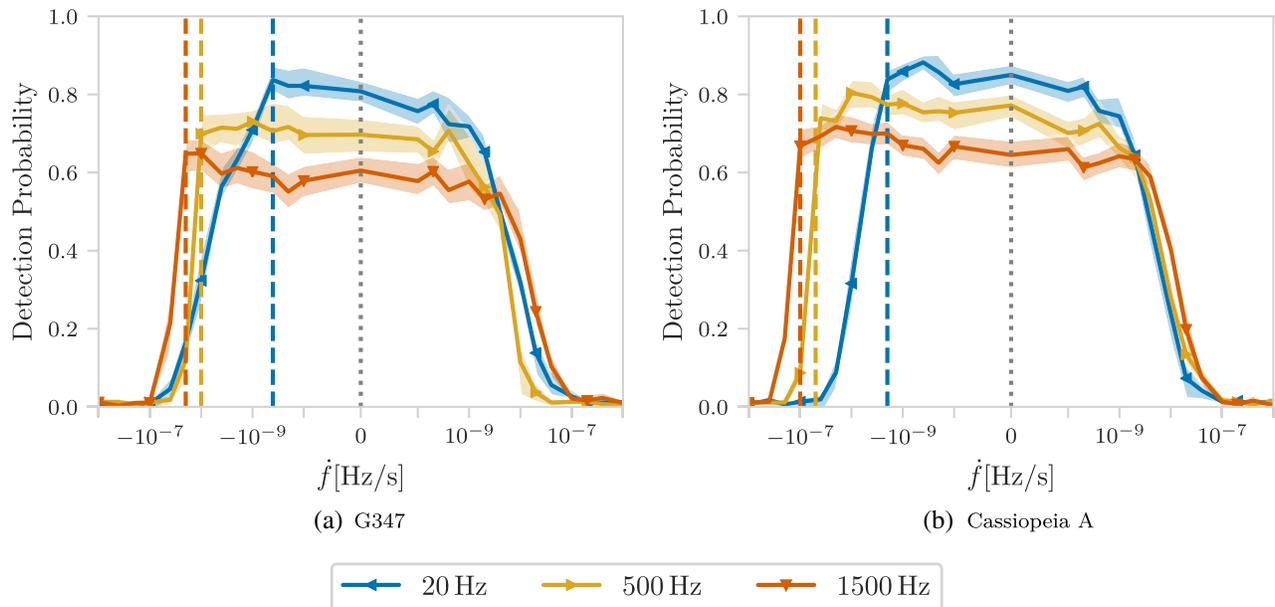

FIG. 4. Detection probability $p_{\rm det}$ versus injected spin-down $\dot{f}$ for networks trained at different frequencies. The x-axis is plotted as a symmetric logarithm, i.e., logarithmic for the larger negative values, linear for $|\dot{f}| < -10^{-10}$ Hz/s and logarithmic for the larger positive values. The vertical dashed lines mark the minimal spin-down $\dot{f}$ used in the training set. Its absolute value increases with frequency. The maximal used spin-down for all cases is 0 Hz/s (dotted line). The shaded areas around each curve show the 95% error regions.

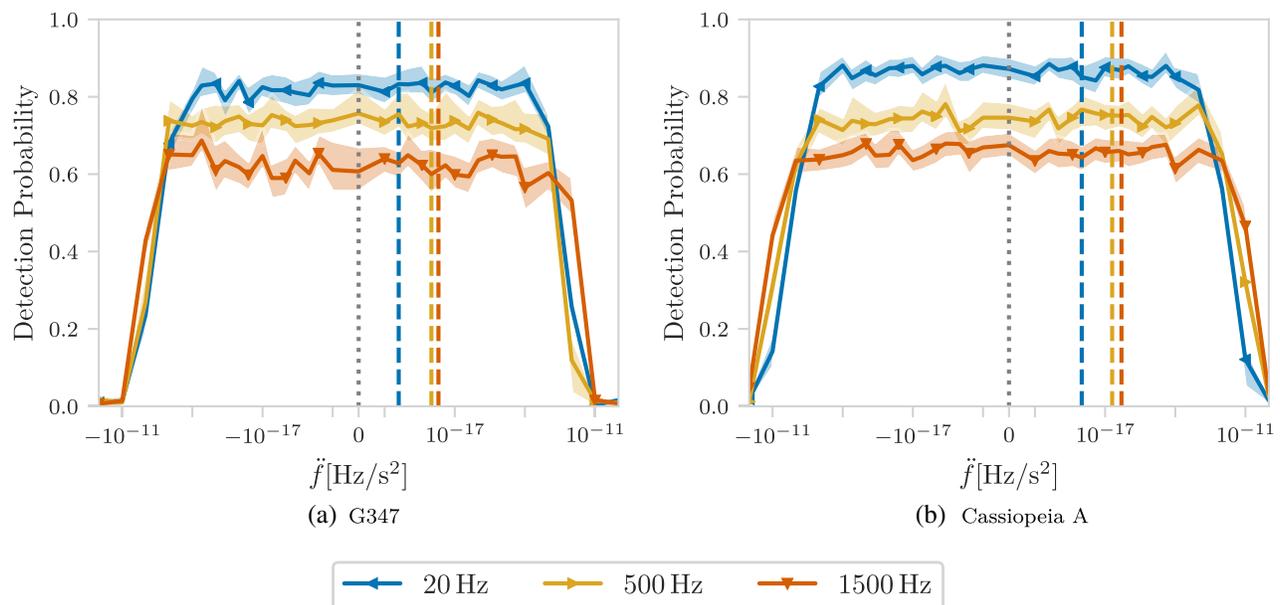

FIG. 5. Detection probability $p_{\rm det}$ versus injected second order spin-down $\ddot{f}$ for networks trained at different frequencies. The x-axis is plotted as a symmetric logarithm, i.e., logarithmic for the larger negative values, linear for $|\ddot{f}| < -10^{-20}$ Hz/s and logarithmic for the larger positive values. The vertical dashed lines mark the maximal second order spin-down $\ddot{f}$ used in the training set, which increases with frequency. The minimal used second order spin-down for all cases is 0 Hz/s² (dotted line). The shaded areas around each curve show the 95% error regions.

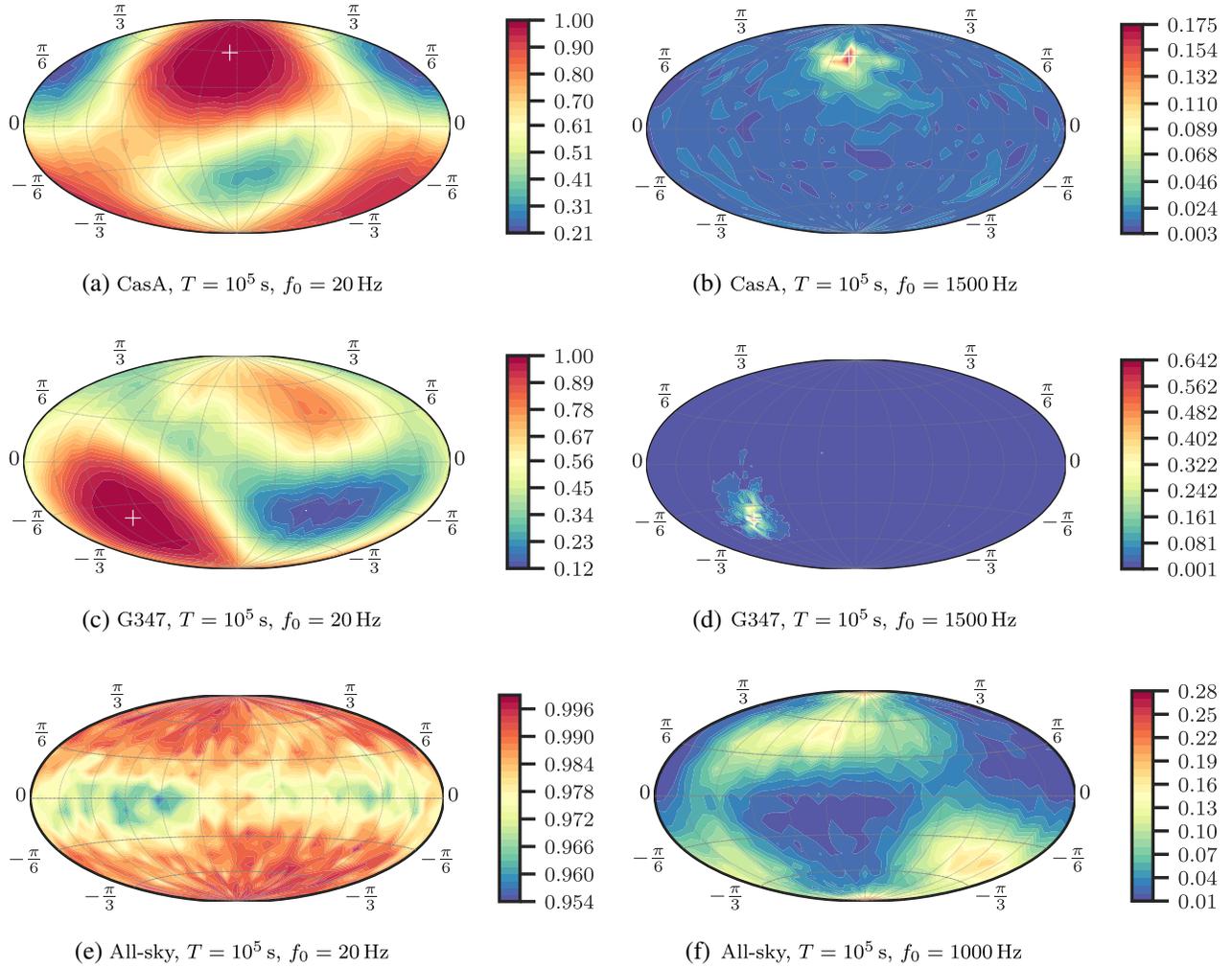

FIG. 6. Detection probability $p_{\text{det}}$ as a function of the sky-position of injected signals in equatorial coordinates (Hammer projection). The detection probability is measured at fixed SNR $\rho = 8.94$. In (a)–(d) the respective sky position of CasA or G347 is marked by a white plus.

The generalization results on the second-order spin-down of the directed searches in Fig. 5 show a qualitatively similar behavior to the first-order spin-down: a plateau of nearly-constant detection probability and a drop starting at about $|\ddot{f}| \gtrsim 10^{-14}$ Hz/s$^2$, approaching zero near $|\ddot{f}| \gtrsim 10^{-11}$ Hz/s$^2$. Contrary to the first-order spin-down results, however, the drop happens many orders of magnitude beyond the trained range of $\ddot{f} \lesssim 10^{-17}$ Hz/s$^2$. This is not surprising, given that a second-order spin-down of this order would only change the signal phase by about $10^{-2}$ rad over the short timespan of $T = 10^5$ s and is therefore still effectively negligible.

### 4. Sky position

Another interesting question is the sensitivity as a function of the sky-position of the signal. For this we measure and plot the DNN detection probability as a function of the sky-position of the signal injections, shown in Fig. 6.

Here we use signals injected at fixed SNR ($\rho = 8.94$) instead of the fixed-depth $\mathcal{D}$ injections used in other tests. By fixing the signal SNR, we can probe the intrinsic sky-position sensitivity of the trained network independently of the detector antenna-patterns while for signals at fixed depth the SNR varies with sky position.

For the directed searches in Fig. 6(a)–6(d) we see a clear preference for the trained sky-position, while sensitivity localization improves with frequency. This is qualitatively similar to how matched filtering behaves, but with a wider sensitive sky region around the targeted sky-position.

For matched filtering we estimate the sensitive region to be of order ∼1 rad at $f = 20$ Hz and ∼$10^{-2}$ rad at $f = 1500$ Hz.

For the all-sky DNNs we see a preference for signals coming from the equatorial poles (latitude $\pm\pi/2$) instead, shown in Figs. 6(e)–6(f). In the $f = 20$ Hz case this effect is relatively small (with a difference of only ∼5% in detection probability), and much more pronounced at $f = 1000$ Hz, where we see some additional structure in right ascension.

We suspect that the observed preference for signals coming from the poles is likely due to their smaller Doppler-broadening compared to signals from the equator, which makes them more concentrated in the frequency domain and therefore easier to "see" for the network. This is also consistent with the DNN detection probability decreasing with increasing signal frequency and increasing observation time, which both result in signals getting more spread out in frequency due to the increase in Doppler broadening.

## V. TESTING NETWORK PERFORMANCE ON REAL DATA

In order to conduct a search for CWs with a DNN, the network has to be able to handle real detector data, which differs in three aspects from the simulated Gaussian data used so far in this study:
(1) potentially different noise levels between detectors
(2) typically noncontiguous data, i.e., gaps in the data due to real gravitational-wave detectors not being in lock continuously
(3) non-Gaussian disturbances in the data, in particular near-monochromatic *lines* that can mimic CWs and trigger false alarms (e.g., see [50] for more discussion).

Here we assess the impact of these effects on the detection performance of a DNN *trained on ideal simulated Gaussian noise without gaps*. In order to separate the different effects, we first test the DNN on simulated Gaussian noise with realistic data gaps, and then with real detector noise, both from a "quiet" undisturbed band and from a disturbed band. The next natural step would be to train networks directly on real detector noise, however this is beyond the scope of this work.

The detector data used is from the LIGO O1 observing run, which can be retrieved from the Gravitational Wave Open Science Center (GWOSC) [51].

### A. Gaussian noise with data gaps

In order to generate data with realistic gaps we randomly select start-times from the LIGO O1 run and retrieve the corresponding gap profile over $T = 10^5$ s. We then generate Gaussian white noise and signals with the same gaps, and we calculate the *duty factor* of this gap profile as $\frac{T_{\text{data}}}{2T} \leq 1$, where $T_{\text{data}}$ is the amount of data from both detectors.

In Fig. 7 we show the results of detection probability as a function of duty factor for two test cases, namely the all-sky benchmarks for $T = 10^5$ s, $f_0 = 20$ Hz and $T = 10^5$ s, $f_0 = 200$ Hz. In both cases we see that the DNN's detection probability (*cross* markers) shows a similar drop in detection probability with decreasing duty factor as matched-filtering does (*solid line*). This indicates that the loss in detection probability stems purely from the intrinsic lowered signal SNR (due to the reduced amount of data), despite the network being trained on fully contiguous data only.

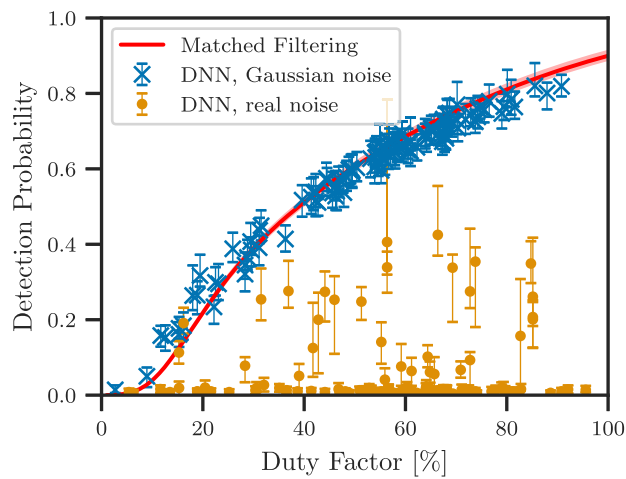
(a) $T = 10^5$ s, $f_0 = 20$ Hz

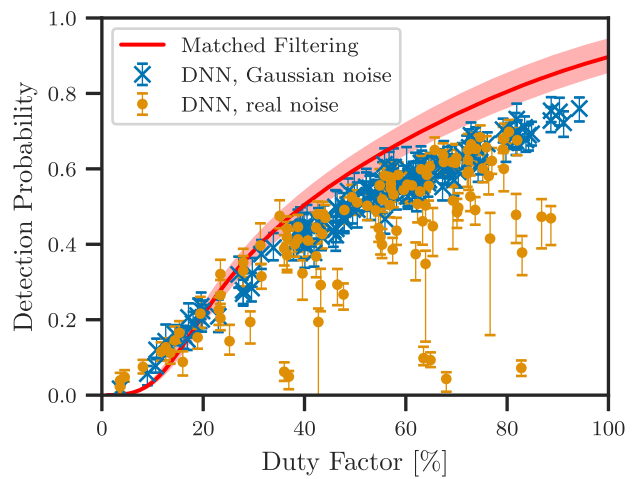
(b) $T = 10^5$ s, $f_0 = 200$ Hz

FIG. 7. Duty factor vs detection probability of an all-sky DNN in Gaussian and real noise. The solid red curve with its shaded region represents the behaviour of matched filtering on Gaussian noise, the blue crosses represent the DNN's performance on Gaussian noise and the yellow circles represent the DNN's performance on real LIGO O1 detector noise. The error bars indicate the 95% confidence interval.

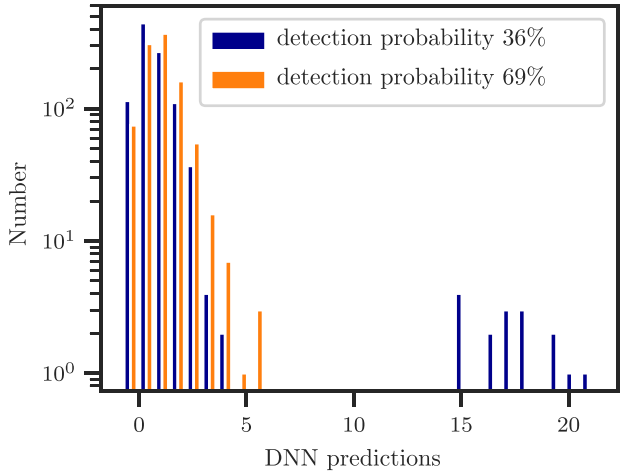

FIG. 8. Histogram of the distribution of DNN detection statistic values (predictions) for 1000 real-noise input samples. The two distributions correspond to two different start-times, with similar duty factors ∼82% and for the same 5 Hz band around 200 Hz. In one case (blue) a disturbance in the data results in a long tail of higher statistic values, which leads to a higher detection threshold at fixed false-alarm, thereby reducing detection probability compared to the undisturbed case (orange).

### B. Performance on real detector data

For this test we use real strain data from the LIGO O1 observing run, with gating and cleaning applied for a recent Einstein@Home[3] search [16].

For a time-span of $T = 10^5$ s with randomly selected start-time during O1, we draw 1000 random 50 mHz-frequency bands from within a 5 Hz band around the test frequency. Using these data samples we determine the detection probability in the usual way: apply the DNN to the data samples (assumed to be pure noise) to determine the detection threshold at $p_{\mathrm{fa}} = 1\%$, then repeat the procedure with added signals of depth $\mathcal{D}_{\mathrm{MF}}^{90\%}$ in order to determine the detection probability (i.e., the fraction of signal samples where the DNN prediction exceeds the threshold).

We found that performing an additional prenormalization of the data by the individual detector noise floors improves the DNN detection performance in the presence of differing noise floors between the two detectors.

The results of the real-data tests are shown in Fig. 7 (filled circles), plotted again as a function of duty factor. For the frequency band at $f = 200$ Hz in Fig. 7 there are many data points basically matching the Gaussian-noise performance, while for others there is a substantial loss in detection probability. This loss can be traced to the presence of "line" disturbances in the data as the disturbed bands create a longer-tailed distribution of DNN detection statistic values for noise inputs. This is illustrated in Fig. 8 for one example. In the low-frequency $f = 20$ Hz case in Fig. 7(a) we see a significant overall drop in detection probability, due to a large number of lines and other disturbances we observed in this frequency band.

## VI. DISCUSSION

In this work we demonstrated that the already-established ability of a deep neural network to search for continuous gravitational waves in the data of a single detector can be extended to two-detector searches. While the larger size of the input data increases the challenge for the DNN the results for short data spans remain reasonably competitive with matched filtering.

On the other hand our architecture searches did not yet yield a reasonably competitive neural network for the longer $T = 10^6$ s data span. Therefore we mostly focused on characterizing the performance of the $T = 10^5$ s networks for now.

Also note that compared to state-of-the-art CW searches the DNN sensitivity achieved here is not yet competitive. For example all-sky searches roughly achieve a sensitivity depth of 30–50 $\mathrm{Hz}^{-1/2}$ (e.g., see [8]) while directed searches go up to 54–83 $\mathrm{Hz}^{-1/2}$ [16].

As was shown in [29] the computing cost of a neural network search is dominated by the training time and the time of a matched-filtering follow-up.[4] This implies that multiple reuses of a trained DNN do not significantly increase the overall computing cost. Training, executing and following-up the $T = 10^5$ s search, using the networks presented in this paper, is roughly two times faster than the respective matched-filtering search.

Furthermore we studied the features of a DNN search directed at a specific sky-position. These directed searches show comparable performance to the all-sky searches at $T = 10^5$ s with respect to the respective matched filter sensitivities, but show less generalization in frequency and first-order spin-down.

A common trend observed here, consistent with the previous study [29], is that the network performance seems to degrade when signals are spread over a wider frequency band, i.e., for higher frequencies, sky positions with more Doppler spreading, and for longer time spans. This shows that the networks still have difficulties learning this aspect of input signals.

We have further shown that DNNs seem relatively robust toward data gaps that differ from the training set, and we found that the impact of unequal detector noise floors can be alleviated by per-detector data normalization. Furthermore, as expected, we find that the performance on real detector noise is significantly reduced in the presence of non-Gaussian disturbances, i.e., "lines".

---

[3]https://einsteinathome.org

[4]The matched-filtering follow-up is currently necessary for a fair comparison as otherwise a matched-filtering search would deliver far more information about candidates than the DNN.

We can identify the following remaining steps toward a competitive and practical DNN search method:

(1) Train the networks on real detector data in order to "learn" to classify disturbances as noise.
(2) Further optimize network architecture to further close the gap to matched filtering under data ideal conditions.
(3) Design a "semicoherent"-type search method by combining the DNN predictions from short time spans (such as $T = 10^5$ s).

## ACKNOWLEDGMENTS

We thank Benjamin Steltner and the AEI CW group for helpful discussion and comments, and for providing the gated and cleaned O1 data. We further thank Marlin Schäfer for valuable input regarding DNNs. All WEAVE Monte-Carlo simulations and DNN training and testing were performed on the ATLAS computing cluster of the Albert-Einstein Institute in Hannover. This research has made use of data, software and/or web tools obtained from the Gravitational Wave Open Science Center (https://www.gw-openscience.org), a service of LIGO Laboratory, the LIGO Scientific Collaboration and the Virgo Collaboration. LIGO is funded by the U.S. National Science Foundation. Virgo is funded by the French Centre National de Recherche Scientifique (CNRS), the Italian Istituto Nazionale della Fisica Nucleare (INFN) and the Dutch Nikhef, with contributions by Polish and Hungarian institutes.